\documentclass[paper]{JHEP3}

\usepackage{latexsym}
\usepackage{amsmath}
\usepackage{amscd}
\usepackage{amsfonts}
\usepackage{amsthm}
\usepackage{graphicx}
\allowdisplaybreaks[1]
\def\half{{\textstyle {1 \over 2}}}
\def\quart{{\textstyle {1 \over 4}}}
\def\noverm#1#2{{\textstyle {#1 \over #2}}}

\def\sgn{{\rm sgn}}

\def\makeatletter{\catcode`\@=11}
\makeatletter
\def\mathbox#1{\hbox{$\m@th#1$}}%
\def\math@ccstyles#1#2#3#4#5#6#7{{\leavevmode
      \setbox0\mathbox{#6#7}%
      \setbox2\mathbox{#4#5}%
      \dimen@ #3%
      \baselineskip\z@\lineskiplimit#1\lineskip\z@
      \vbox{\ialign{##\crcr
             \hfil \kern #2\box2 \hfil\crcr
             \noalign{\kern\dimen@}%
             \hfil\box0\hfil\crcr}}}}
\def\mathaccstyles{\math@ccstyles\maxdimen}
\def\maththroughstyles{\math@ccstyles{-\maxdimen}}
\def\unitmatrixDT%
 {\maththroughstyles{.45\ht0}\z@\displaystyle {\mathchar"006C}\displaystyle 1}
\def\leftrightarrowfill{$\mathsurround0pt \mathord\leftarrow
       \mkern-6mu\cleaders\hbox{$\mkern-2mu \mathord- \mkern-2mu$}\hfill
       \mkern-6mu \mathord\rightarrow$}

\def\overleftrightarrow#1{\vbox{\ialign{##\crcr
        \leftrightarrowfill\crcr\noalign{\kern-1pt\nointerlineskip}%
        $\hfil\displaystyle{#1}\hfil$\crcr}}}

\def\unitm#1{\unitmatrixDT_{#1}}

\title{De Sitter solutions in $N=4$ matter coupled supergravity}
\author{M.~de Roo and D.~B.~Westra,\\
Institute for Theoretical Physics\\
   Nijenborgh 4, 9747 AG Groningen,
     The Netherlands\\
     E-mail: \email{m.de.roo@phys.rug.nl\\ d.b.webstra@phys.rug.nl}}
\author{S.~Panda\\
    Harish-Chandra Research Institute \\
    Chatnag Road, Jhusi, Allahabad 211019, India\\
        E-mail: \email{panda@mri.ernet.in}}

\preprint{UG-02/44\\ MRI-P-021102\\ \hepth{0212216}}

\abstract{We investigate the scalar potential of gauged $N=4$
 supergravity with matter. The extremum in the $SU(1,1)/U(1)$
 scalars is obtained for an arbitrary number of  matter
 multiplets.
 The constraints
 on the matter scalars are solved in terms of an explicit
 parametrisation of an $SO(6,6+n)$ element.
 For the case of six matter multiplets we discuss
 both compact and noncompact gauge groups.
 In an example involving noncompact groups and four scalars
 we find a potential with an absolute minimum and a positive
 cosmological constant.
}


\begin{document}

\section{Introduction\label{Intro}}

The evidence for a positive cosmological constant has led to a
renewed interest in gauged supergravity theories. The presence
of a scalar potential in these theories opens the possibility
of obtaining, at
the extremum of the potential, $V_0$, a nonzero cosmological constant.
In the past the interest in this field was concentrated on zero or negative
values of $V_0$, in view of preserving some of the supersymmetries for
phenomenological applications\footnote{For recent work in this direction
see, e.g., \cite{AFV,AFLV}.}.
For positive
values of $V_0$, or de Sitter solutions, supersymmetry is necessarily
completely broken.

In this paper we will investigate properties of gauged $N=4$ supergravity
in four dimensions with positive $V_0$. Gauged supergravity itself contains
two physical scalars, which take values in the coset $SU(1,1)/U(1)$.
The potential due to these scalars has been
investigated in great detail in the
past \cite{DFR}-\cite{PW1}. Here we want to extend this work to
the case where an arbitrary number of matter multiplets is included.

There are a number of obstructions to the existence of solutions
in four-dimensional supergravities with a positive cosmological constant
\cite{GWG,dWHDS,MN}. These obstructions show up in particular in theories
obtained by
dimensional reduction from eleven or ten dimensions. However, in
four dimensions much more is possible. In $N=4$ supergravity one can for
instance add additional matter beyond what would be expected from
string theory, one can gauge some of the global symmetries of
the supergravity theory, and one can introduce additional parameters
\cite{MdRPW1}, called $SU(1,1)$ angles, which give the
matter multiplets  different $SU(1,1)$ orientations. Although gauged
supergravities may be  related to Scherk-Schwarz
reductions of higher dimensional theories, it is then still
not clear how to introduce the $SU(1,1)$ angles.
For $N=2$ theories a recent investigation \cite{FTVP}
revealed that it
is possible to obtain stable de Sitter vacua. The analogue of the
$N=4$ $SU(1,1)$ angles played a crucial role in
\cite{FTVP}.

An interesting investigation of scalar potentials in extended
supergravity theories was performed in \cite{KLPS}. These authors also
consider $N=4$, but without additional matter. They
remark that in all cases considered there is a simple
relation between the value of the potential at its extremum and the
masses of the scalar excitations at the extremum:
\begin{equation}
\label{Vrel}
   V_0 \simeq {\partial^2 V\over \partial x^2}\bigg|_0\,.
\end{equation}
This relation makes these examples less suitable for application in
the context of inflationary models, although in some scenarios
they do not seem to be excluded (see \cite{KLPS} for a discussion on
this matter). In Section \ref{sugra} we will make this relationship
more precise for the $SU(1,1)$ scalars.

In this paper we further develop the formalism of gauged
$N=4$ supergravity with matter, and give examples with positive extrema
of the potential.
In the remainder of this
Introduction we will present some basics of gauged $N=4$ supergravity.
The analysis of the dependence on the $SU(1,1)$ scalars is done
in Section \ref{sugra}.  The matter scalars are considered
in Section \ref{matter}. We solve the constraint for these scalars, and
discuss properties of the potential for general matter fields.
In Sections \ref{examples}
we work out a number of explicit examples for gauge groups
$SO(3)^4$, $SO(3)^2\times SO(2,1)^2$, and $SO(2,1)^4$.
In Section \ref{discussion}
we will discuss some additional issues and further work.

We consider gauged $N=4$ supergravity coupled to $n$ vector multiplets.
The bosonic part of the Lagrangian density reads
\cite{MdRPW1}
\footnote{The indices $\alpha,\beta,\ldots$ take on values $1$ and $2$,
indices $R,S\ldots$ the values $1,\ldots, 6+n$, and
the indices $a,b,\ldots$ the values $1,\ldots,6$. The metric $\eta_{RS}$ can
be chosen as diag$(-1,-1,-1,-1,-1,-1,+1,\ldots,+1)$, with $n$ positive
entries. In comparison to \cite{MdRPW1} we have replaced
the complex scalars $\phi_{ij}{}^R$ by real scalars $Z_a{}^R$:
$\phi_{ij}{}^R = \half Z_a{}^R (G^a)_{ij}$, where the $G^a$ are
 six matrices which ensure that $Z_a{}^R$ transforms as a
vector under $SO(6)$. This redefinition is given (in a slightly different
normalisation) in \cite{MdR1}.}:
\begin{eqnarray}
\label{action}
   e^{-1}{\cal L} &=& -\half R
  +\half \big( \partial_\mu\phi^\alpha\partial^\mu\phi_\alpha
  + \phi^\alpha\partial_\mu\phi_\alpha \phi^\beta\partial^\mu\phi_\beta\big)
  \nonumber\\
  && -\half\eta_{RS}\partial_\mu Z_a{}^R\partial^\mu Z_a{}^S
     -\noverm{1}{8}\eta_{RS}\eta_{TU}
    Z_a{}^R \overleftrightarrow{\partial_\mu}Z_b{}^S\,
    Z_a{}^T \overleftrightarrow{\partial^\mu}Z_b{}^U
  \nonumber\\
  && -V(\phi,Z)
  \nonumber\\
  &&  +\eta_{RS}\big(
      -\quart F_{\mu\nu}^{+\,R} F^{\mu\nu\,+\,S}
      {1\over \Phi_{(R)}}(\phi^1_{(R)}-\phi^2_{(R)})
  \nonumber\\
  && -{1\over \Phi_{(R)}}F_{\mu\nu}^{+\,R}K^{\mu\nu\,+\,S}
    - \half {\Phi^*_{(R)}\over \Phi_{(R)}}
        K_{\mu\nu}^{+\,R}K^{\mu\nu\,+\,S} + {\rm h.c.} \big)\,.
\end{eqnarray}
The scalars $\phi_\alpha$ ($\phi^1=(\phi_1)^*, \phi^2=-(\phi_2)^*$)
transform under global $SU(1,1)$ and
local $U(1)$, the $Z_a{}^R$ transform under local $SO(6)\times SO(n)$, and
under global $SO(6,n)$. The scalars satisfy the
constraints
\begin{eqnarray}
\label{conphi}
   &&\phi^\alpha\phi_\alpha = 1\,,
   \\
\label{conZ}
   &&\eta_{RS}Z_a{}^R Z_b{}^S = -\delta_{ab}\,.
\end{eqnarray}
Due to these constraints and the local symmetry the scalars are restricted to
cosets $SU(1,1)/U(1)$ (two physical scalars) and $SO(6,n)/SO(6)\times SO(n)$
($6n$ physical scalars).

There is a certain freedom in coupling the vector multiplets: for each
multiplet we can introduce an $SU(1,1)$ element, of which only a
single angle $\alpha$ turns out to be important. These angles $\alpha_R$ appear in
the kinetic terms of the vectors in the form
\begin{equation}
    \phi^1_{(R)} = e^{i\alpha_R}\phi^1\,,\quad
    \phi^2_{(R)} = e^{-i\alpha_R}\phi^2\,,\quad
    \Phi_{(R)} = e^{i\alpha_R}\phi^1 + e^{-i\alpha_R}\phi^2\,.
\end{equation}
The gauge group has to be a subgroup of $SO(6,n)$.
For a semi-simple gauge group the $\alpha_R$ have to be the same for all
$R$ belonging to the same factor of the gauge group. The gauging
breaks the global $SO(6,n)$ symmetry of the ungauged theory.

In (\ref{action}) we have made explicit the dependent gauge fields of
the local $U(1)$ and $SO(6)$ symmetries. The kinetic terms for the vectors
still contain the auxiliary field $T_{\mu\nu}{}^{ij}$, which must be
eliminated by solving its equation of motion. Here it is again useful to go
to a real basis and to define $T_{\mu\nu}{}^a = \half
T_{\mu\nu}{}^{ij}G^a{}_{ij}$.
In this form we have  $K_{\mu\nu}{}^R=T_{\mu\nu}{}^a Z_a{}^R$. The
equation of motion for $T$ is
\begin{equation}
  T_{\mu\nu}{}^a Q_{ab} + \eta_{RS}{1\over \Phi_{(R)}}F_{\mu\nu}^{+\,R}
    Z_b{}^S = 0\,,
\end{equation}
with
\begin{equation}
   Q_{ab} = \eta_{RS}{\Phi^*_{(R)}\over \Phi_{(R)}}Z_a{}^R Z_b{}^S\,.
\end{equation}
The vector kinetic terms can then be written as:
\begin{eqnarray}
   {\cal L}_{{\rm vector}} &=&
    -\quart\eta_{RS}
       F_{\mu\nu}^{+R} F^{\mu\nu\,+\,S}
      {1\over \Phi_{(R)}}(\phi^1_{(R)}-\phi^2_{(R)}) +
  \nonumber\\
  && + \half\eta_{RS}\eta_{TU}{1\over \Phi_{(R)}}F^{\mu\nu\,+\,R}
             Z_a{}^S (Q^{-1})_{ab} Z_b{}^T
      {1\over \Phi_{(T)}}F_{\mu\nu}^{+\,U} + {\rm h.c.}\,.
\end{eqnarray}
This simplifies in  two special cases: firstly, if all
$SU(1,1)$ angles are equal
($\Phi_{(R)}=\Phi$) we have $Q_{ab}=-\delta_{ab}\Phi^*/\Phi$,
secondly, with six vector multiplets (\ref{conZ}) can be solved by
$Z_a{}^R=\delta_a{}^R$, and $Q_{ab} = -\delta_{ab}\Phi_{(a)}^*/\Phi_{(a)}$.

The scalar potential reads
\begin{equation}
  V(\phi,Z) = \big(\quart Z^{RU}Z^{SV}(\eta^{TW}+\noverm{2}{3}Z^{TW})
    -\noverm{i}{36}Z^{RSTUVW}\big)\Phi^*_{(R)}f_{RST}\Phi_{(U)}f_{UVW}\,,
\end{equation}
where $Z^{RS}=Z_a{}^RZ_a{}^S$ and
$Z^{RSTUVW}=\epsilon^{abcdef}Z_a{}^RZ_b{}^SZ_c{}^TZ_d{}^U Z_e{}^V Z_f{}^W$.
The structure constants $f_{RST} = f_{RS}{}^T\eta_{VT}$ are totally
antisymmetric.

\section{The scalar potential: the $SU(1,1)/U(1)$ scalars\label{sugra}}

In this section we will discuss properties of the scalar potential that are
independent of the specific matter content and choice of gauge group.
The potential can be written in the form:
\begin{equation}
\label{pot}
   V = \sum_{i,j}\, (R^{(ij)}\, V_{ij} + I^{(ij)} \,W_{ij})\,.
\end{equation}
The indices $i,j,\ldots$ label the different factors in the gauge group
$G$, which we will take to be semi-simple.
$R$ and $I$ contain the $SU(1,1)$ scalars and depend on the
gauge coupling constants and the $SU(1,1)$ angles,
$V$ and $W$ contain the
structure constants, depend on the matter
fields, and are symmetric resp.\ anti-symmetric
in the indices $i,j$.
We have:
\begin{eqnarray}
  R^{(ij)} &=& {g_i g_j\over 2}  (\Phi^*_i\Phi_j + \Phi^*_j\Phi_i)\nonumber\\
           &=& g_ig_j\left(\cos(\alpha_i-\alpha_j){1+r^2\over 1-r^2}
                - {2r\over 1-r^2}\cos(\alpha_i+\alpha_j+\varphi)
                \right)\,,\\
  I^{(ij)} &=& {g_ig_j \over 2i} (\Phi^*_i\Phi_j - \Phi^*_j\Phi_i)\nonumber\\
           &=& -g_ig_j\sin(\alpha_i-\alpha_j)\,.
\end{eqnarray}
The fields $r$ and $\varphi$ represent the scalars of the $SU(1,1)/U(1)$ coset,
we have solved the constraint (\ref{conphi}) in a suitable $U(1)$ gauge as
\begin{equation}
   \phi_1 = {1\over \sqrt{1-r^2}}\,,\qquad
   \phi_2 = {re^{i\varphi}\over \sqrt{1-r^2}}\,.
\end{equation}

In this section we will discuss the extremum of this potential
for $r$ and $\varphi$.
We introduce the quantities
\begin{eqnarray}
\label{CS}
  C_{\pm} &=& \sum_{ij} g_ig_j\cos(\alpha_i\pm\alpha_j)V_{ij}\,,\quad
  S_{+} = \sum_{ij} g_ig_j\sin(\alpha_i+\alpha_j)V_{ij} \,,\\
\label{Tmin}
  T_{-}   &=& \sum_{ij} g_ig_j\sin(\alpha_i-\alpha_j)W_{ij} \,,
\end{eqnarray}
and write the potential as
\begin{equation}
\label{potrphi}
   V = C_-\,{1+r^2\over 1-r^2} - {2r\over 1-r^2}\,
     \big(C_+\cos\varphi - S_+\sin\varphi\big) - T_-\,.
\end{equation}
One finds that the extremum in $\varphi$ is obtained for
\begin{equation}
  \cos\varphi_0 = {s_1 C_+\over \sqrt{C_+^2 + S_+^2}}\,,\quad
  \sin\varphi_0 = -{s_1 S_+\over \sqrt{C_+^2 + S_+^2}}\,,\ (s_1=\pm 1)\,.
\end{equation}
The equation for the extremum in $r$ becomes, for $\varphi=\varphi_0$,
\begin{equation}
\label{reom}
  r^2 + 1 -s_1 {2rC_-\over \sqrt{C_+^2 + S_+^2}} = 0\,,
\end{equation}
which, for
\begin{equation}
\label{condition}
 \Delta \equiv C_-^2-C_+^2-S_+^2>0\,,
\end{equation}
is solved by ($s_2=\pm 1$, a priori independent of $s_1$):
\begin{equation}
  r_0 = {1\over \sqrt{C_+^2 + S_+^2}}
    \left( s_1 C_- + s_2 \sqrt{C_-^2- C_+^2 - S_+^2}\right)\,.
\end{equation}
Now consider the signs $s_1$ and $s_2$.
 The condition
$r_0<1$ leads to
\begin{equation}
  s_1s_2 C_- +   \sqrt{C_-^2- C_+^2 - S_+^2} < 0\,,
\end{equation}
which implies $s_1s_2= - \sgn C_-$. To have $r_0 \ge 0$ we need
$s_1 = \sgn C_-$, so that $s_2=-1$.
After substitution of $r_0$ and $\varphi_0$ in $V$, we obtain
\begin{equation}
\label{pot1}
  V_0 = \sgn{C_-}\,\sqrt{C_-^2- C_+^2 - S_+^2}  - T_- \,.
\end{equation}

In the case that all $SU(1,1)$ angles $\alpha_i$
vanish,  $S_+=T_-=0$ and $C_-=C_+$, and (\ref{reom}) leads
to $r_0=1$, which is a singular point of the parametrisation:
there is no extremum.  This is
the generalisation of the Freedman-Schwarz potential \cite{FS}
to the case of general matter coupling. The sign of the potential
is the sign of $C_-$. We will not discuss
this situation any further.

If $\Delta>0$, which requires some of the $SU(1,1)$ angles to be
different,
the extremum (\ref{pot1}) exists and can be further simplified
by looking in more detail at $C_{\pm},\ S_{+}$. We find
\begin{eqnarray}
  &&{C_-^2- C_+^2 - S_+^2} = \sum_{ij}\sum_{kl}
   g_ig_jg_kg_l\,V_{ij}V_{kl}
  (\cos(\alpha_i-\alpha_j)\cos(\alpha_k-\alpha_l) \nonumber\\
  &&\qquad\qquad
  - \cos(\alpha_i+\alpha_j)\cos(\alpha_k+\alpha_l)
  - \sin(\alpha_i+\alpha_j)\sin(\alpha_k+\alpha_l))\nonumber\\
\label{Delta}
  &&\quad =\ 2\,\sum_{ij}\sum_{kl} g_ig_jg_kg_l\,
   V_{ij}V_{kl} \sin(\alpha_i-\alpha_k)\sin(\alpha_j-\alpha_l)\,.
\end{eqnarray}
We see that
the potential at the extremum in $r,\varphi$ depends only on the combinations
$g_i g_j \sin(\alpha_i-\alpha_j)$.
This was known for gauged supergravity
without additional matter, and now turns out to be a general property.

If the condition (\ref{condition}) holds we find an extremum for
the $SU(1,1)$ scalars. To see what happens at the extremum
it is useful to work
out the second derivatives of the potential. We will do this
with respect to the variables
\begin{equation}
    x=r\cos\varphi\,,\qquad y=r\sin\varphi\,.
\end{equation}
One easily finds that in the extremum
\begin{eqnarray}
\label{second}
   {\partial^2 V\over \partial x^2}\bigg|_{0}
   &=&  {\partial^2V\over \partial y^2}\bigg|_{0} =
   {4 \over (1-r_0^2)^2}\, \sgn C_- \sqrt{C_-^2-C_+^2-S_+^2}\,,\quad
  {\partial^2V\over \partial x\partial y}\bigg|_{0} = 0\,.
\end{eqnarray}
If $\sgn C_-\,>\,0\,(<0)$ the potential has a minimum (maximum)
for the scalars $x,y$.
We see that, up to a scale factor, the second derivatives are equal
to the first term in the potential at the extremum (\ref{pot1}).
The scale factor can be understood by considering the kinetic term
for the scalars. It can be read off from (\ref{action}), and, after
expressing the action in terms of the variables $x,y$ we find:
\begin{equation}
    {\cal L}_{{\rm kin,\phi}} = -{2\over (1-r^2)^2}
    \bigg( \partial_\mu x\partial^\mu x + \partial_\mu y \partial^\mu y\bigg)\,.
\end{equation}
To have a proper normalisation for these fields requires a rescaling:
\begin{equation}
    x = \half x' (1-r_0)^2\,,\qquad  y = \half y' (1-r_0)^2\,.
\end{equation}
The contribution to the action of the new scalars $x',y'$, including the
potential, then looks as follows:
\begin{eqnarray}
   {\cal L}_{\phi} &=&
   - \half \left( {1-r_0^2\over 1-r^2}\right)^2
    \big( \partial_\mu x'\partial^\mu x' + \partial_\mu y' \partial^\mu y'\big)
   - \sgn{C_-}\,\sqrt{C_-^2- C_+^2 - S_+^2}   + T_-
  \nonumber\\
\label{actionphi}
  && -\half\, \sgn{C_-}\,\sqrt{C_-^2- C_+^2 - S_+^2}\,\big( x^{\prime\,2}
    + y^{\prime\,2}\big)\ +\ \ldots\,.
\end{eqnarray}
We find that in these variables the equality $V_0 = m^2$ is exact, except
for the term $T_-$ (\ref{Tmin}). The reason is of course
that $T_-$ only depends on the matter scalars
and therefore does not contribute to the second derivatives. The
presence of $T_-$ therefore violates the equality (\ref{Vrel}).

In (\ref{actionphi}) we also see that the (mass)${}^2$ is determined by the
sign of $C_-$. The sign of $C_-$ also determines the sign of the first term
in the extremum of the potential. If we go back to the potential itself
(\ref{potrphi}), we see that for $r\to 1$ the potential behaves as
\begin{equation}
   V \to {1\over 1-r} \big( C_- - C_+\cos\varphi + S_+\sin\varphi \big)\,.
\end{equation}
Therefore along the unit circle the potential goes to infinity, and the
sign in this limit, assuming $\Delta>0$, is again the same
as $\sgn C_-$. In the origin $r\to 0$ the potential and its derivatives
are well behaved.

We conclude that if (\ref{condition}) holds and if $\sgn C_->0$,
the potential as a function of the
$SU(1,1)$ scalars has a minimum, and goes to $+\infty$ along the unit circle.
The value of the potential gets a contribution which has the same sign as
$C_-$, but the value and sign of the minimum depends also on $T_-$.
In the next
sections we will try to find examples with this behaviour.

\section{The scalar potential: matter multiplets\label{matter}}

The $N=4$ vector multiplet contains one vector and six scalar fields,
and, in the
construction of $N=4$ Poincar\'e supergravity with the superconformal
method,
six multiplets are required to gaugefix the superconformal
symmetries. Adding $n+6$ vector multiplets to conformal
supergravity thus gives $n$ matter multiplets, and $6n$
physical scalars. These are however expressed in terms of
$6n+36$ variables (the $Z_a{}^R)$,
of which 36 are eliminated by gaugefixing the local $SO(6)$
symmetry and by the constraint (\ref{conZ}).
The constraint implies that $Z$ corresponds to the first six
rows of an $SO(6,n)$ element.

The constraint is a complication in the analysis of the scalar sector
of the theory. Fortunately, there is a nice way to parametrise the
solution of the constraint,
as was remarked in \cite{BMdR}\footnote{Other parametrisations are
discussed in \cite{PW1}}. The dimensional reduction of the $D=10,\ N=1$
supergravity theory, coupled to $m$ vector multiplets, gives after
reduction $12+m$ vectors in $D=4$. The scalars resulting
from this reduction are the 21 scalars coming from the $D=10$ metric, the
15 scalars coming from the $D=10$ two-form, and $6m$ scalars from the
$D=10$ vector multiplets, altogether $6m+36$.
These scalars parametrise the coset
$SO(6,6+m)/SO(6)\times SO(6+m)$ \cite{MS}:
\begin{eqnarray}
\label{Nmatrix}
N&=& \left(\begin{matrix} G^{-1} & G^{-1} (B+W) &
  \sqrt{2}G^{-1} U\cr
  (-B+W)G^{-1} & (G-B+W)G^{-1} (G+B+W) & \sqrt{2}
  (G-B+W)G^{-1} U \cr
  \sqrt{2} U^T G^{-1} & \sqrt{2} U^T G^{-1} (G+B+W)
  & \unitm{m} + 2 U^T G^{-1} U \cr \end{matrix}\right),
\end{eqnarray}
where $G(B)$ are (anti)-symmetric $6\times 6$ matrices, $U$
is a $6\times m$ matrix, and $W=U^TU$. The matrix $N$
satisfies
\begin{equation}
  N\gamma N^T = \gamma\,,
\end{equation}
with
\begin{equation}
 \gamma = \left(\begin{matrix} 0 & \unitm{6}&0\cr
  \unitm{6}&0&0\cr0&0& \unitm{m}\cr\end{matrix}\right) \,.
\end{equation}
We now transform the metric $\gamma$ to the metric $\eta$,
and $N$ to $N^\prime$ by
\begin{equation}
  \eta = M\gamma M^T\,,\quad N^\prime = M N M^T\,,
\end{equation}
with
\begin{equation}
 M = {1\over \sqrt{2}}\left(\begin{matrix} \unitm{6} & -\unitm{6}&0\cr
  -\unitm{6}& -\unitm{6}&0\cr0&0&\unitm{m}\\\end{matrix}\right) \,.
\end{equation}
From the matrix $N^\prime$ we can read off what $Z_a^R$ is: the
first six rows $N^\prime$. Thus we have an explicit parametrisation
of the $Z_a{}^R$ for $n=6+m$.

We emphasize however that this does not imply that the gauged $D=4$
theory with arbitrary $SU(1,1)$ angles follows by reduction
from $D=10$. We only use the
reduction from $D=10$ to solve the constraints.

In this paper we will limit ourselves to the case where $m=0$,
or $U=0$ in (\ref{Nmatrix}). This corresponds to six
vector multiplets added to Poincar\'e supergravity. We split the
indices $R,S,\ldots$ of $\eta_{RS}$ in $A,B,\ldots=1,\ldots,6$,
($\eta_{AB}=-\delta_{AB}$) and $I,J,\ldots=7,\ldots,12$,
($\eta_{IJ}=+\delta_{IJ}$). The scalar constraint (\ref{conZ})
then reads
\begin{equation}
   X X^T - Y Y^T = \unitm{6}\,,
\end{equation}
where $X_{a}{}^A = Z_a{}^A,\ Y_A{}^I = Z_a{}^I$. $X$ and $Y$ are
both $6\times 6$ matrices, which together form the first
six rows of the matrix $N^\prime$:
\begin{eqnarray}
  X &=& {1\over 2}\left(
           G + G^{-1} + BG^{-1} - G^{-1}B - B G^{-1} B
      \right)\,, \\
  Y &=& {1\over 2}\left(
           G - G^{-1} - BG^{-1} - G^{-1}B - B G^{-1} B
       \right)\,.
\end{eqnarray}
The scalar potential will depend on $Z^{AB}=(X^TX)^{AB},
\ Z^{AI}=(X^TY)^{AI}$ and $Z^{IJ}=(Y^TY)^{IJ}$.

In the examples in Section \ref{examples}
we will further simplify matters by choosing for $G$ and $B$:
\begin{equation}
\label{matterpar}
  G = \left(\begin{matrix}
            a \unitm{3} & 0 \\
            0  & a \unitm{3}
        \end{matrix}\right)\quad (a>0)\,,\qquad
  B = \left(\begin{matrix}
            0 & b \unitm{3} \\
        -b \unitm{3} & 0
        \end{matrix}\right) \,,
\end{equation}
which gives
\begin{equation}
  X = {a^2+b^2+1\over 2a} \unitm{6}\,,\quad
  Y = {1\over 2a} \left(
         \begin{matrix}
              (a^2+b^2-1)\unitm{3} & -2b\unitm{3} \\
                 2b\unitm{3} &  (a^2+b^2-1)\unitm{3}
          \end{matrix}
      \right)\,.
\end{equation}
The variables $Z^{RS}$ are then easily determined to be:
\begin{eqnarray}
  Z^{AB} &=& { (a^2+b^2+1)^2  \over 4a^2} \unitm{6} \,,
  \nonumber\\
  Z^{AI} &=& {a^2+b^2+1\over{4a^2}}
    \left(\begin{matrix}
       (a^2+b^2-1)\unitm{3} & -2b\unitm{3} \\
       2b\unitm{3} & (a^2+b^2-1)\unitm{3}
    \end{matrix}\right) \,,
  \nonumber\\
  Z^{IJ} &=& {1\over 4a^2}\bigg( (a^2+b^2+1)^2 - 4a^2 \bigg)\unitm{6} \,.
\end{eqnarray}

\section{Examples\label{examples}}

With the ingredients of Sections \ref{sugra} and \ref{matter} we will now
work out a number of examples. We have seen that the analysis of the
$SU(1,1)$ scalars in Section \ref{sugra} depends crucially on condition
(\ref{condition}), $\Delta>0$.
We will only work out cases for which this condition
is satisfied for all values of the matter fields $a$ and $b$ in our
parametrisation of $G$ and $B$ (\ref{matterpar}). This excludes among
others potentials of the Freedman-Schwarz type \cite{FS}. To evaluate
(\ref{condition}) we will need the contributions $V_{ij}$ and $W_{ij}$
to the potential, see (\ref{pot}). These are given by
\begin{eqnarray}
\label{Vij}
  V_{ij} &=& \quart Z^{RU}Z^{SV}\,(\eta^{TW} + \noverm{2}{3} Z^{TW})\,
          f^{(i)}{}_{RST} f^{(j)}{}_{UVW}\,,
\\
\label{Wij}
  W_{ij} &=& \noverm{1}{36}\epsilon^{abcdef}
      Z_a{}^RZ_b{}^S Z_c{}^T Z_d{}^U Z_e{}^V Z_f{}^W\,
      f^{(i)}{}_{RST} f^{(j)}{}_{UVW}\,,
\end{eqnarray}
where the $f^{(i)}$ are the structure constants for the different
factors of the semi-simple gauge group.

\subsection{$SO(3)^4$\label{examplesc}}

First we will consider a product of compact groups: $G=SO(3)^4$.
For the group $SO(3)$ the structure constants are
\begin{equation}
   f_{RS}{}^T = -\epsilon_{RST}\,,\quad f_{RST} = f_{RS}{}^U \eta_{UT}\,.
\end{equation}
The sign of $f_{RST}$, which is completely anti-symmetric,
therefore depends on the sign of the element $\eta_{TT}$.
The four $SO(3)$ groups are labelled $i=1,\ldots,4$, and
associated with
the values $R,S,\ldots$ in the following way:
\begin{equation}
R,S,\ldots = \overbrace{\,1\ 2\ 3\,}^{i=1}\ \overbrace{\,4\ 5\ 6\,}^{i=2}\
\overbrace{\,7\ 8\ 9\,}^{i=3}\ \overbrace{\,10\ 11\ 12\,}^{i=4} \,.
\end{equation}
We then obtain
the contributions $V_{ij}$ and $W_{ij}$ to the potential:
\begin{eqnarray}
  V_{11} &=& V_{22}
   = {1\over 64a^6}\,(1+a^2+b^2)^4\big( (1+a^2+b^2)^2 - 6a^2\big)\,,
   \nonumber\\
  V_{33} &=& V_{44}
   = {1\over 64a^6}\,\big((1+a^2+b^2)^2-4a^2\big)^2
       \big((1+a^2+b^2)^2+2a^2)\big)\,,
   \nonumber\\
  V_{12} &=& V_{34} = 0\,,
   \nonumber\\
  V_{13} &=&  V_{24} =
   -{1\over 64a^6}\,(a^2+b^2+1)^3(a^2+b^2-1)^3\,,
   \nonumber\\
  V_{23} &=&  -V_{14} =
   -{1\over 64a^6}\,(a^2+b^2+1)^3 (2b)^3\,,
   \nonumber\\
  W_{12} &=& {1\over 64a^6} \, (1+a^2+b^2)^6\,,
   \nonumber\\
  W_{34} &=& {1\over 64a^6}\, \big(1+a^2+b^2)^2-4a^2\big)^3 \,,
   \nonumber\\
  W_{14} &=&  -W_{23} =
      -{1\over 64a^6} \, (a^2+b^2+1)^3(a^2+b^2-1)^3  \,,
   \nonumber\\
  W_{13} &=&  W_{24} =
         -{1\over 64a^2}\, (a^2+b^2+1)^3 (2b)^3 \,.
\end{eqnarray}
Using these, we will discuss the properties of the potential for
some special cases:
\begin{itemize}
\item $g_3=g_4=0$. We assume $\alpha_1\ne \alpha_2$.
Then $C_-=g_1^2V_{11}+g_2^2V_{22}= V_{11}(g_1^2+g_2^2)$.
$V_{11}$ is negative inside the circle
\begin{equation}
\label{circle1}
  (a-\half\sqrt{6})^2+b^2=\noverm{1}{2}\,.
\end{equation}
We find
\begin{equation}
   \Delta = 4V_{11}^2g_1^2g_2^2\sin^2(\alpha_1-\alpha_2)
\end{equation}
which is never negative: the extremum for the $SU(1,1)$ scalars discussed
in Section \ref{sugra} exists for all values of $a$ and $b$ if $\Delta>0$,
on the circle (\ref{circle1})
$V_{11}=C_{\pm}=S_+=\Delta=0$ and
the potential (\ref{potrphi})is independent
of $r$ and $\varphi$.
The matter potential in the extremum of the $SU(1,1)$ scalars reads
\begin{equation}
   V =  V_{11} | 2g_1g_2\sin(\alpha_1-\alpha_2)|
         - 2g_1g_2\sin(\alpha_1-\alpha_2) W_{12}\,.
\end{equation}
Consider the case $g_1g_2\sin(\alpha_1-\alpha_2)\,<\,0$.
The matter potential then reads:
\begin{eqnarray}
  V &=& 2|g_1g_2\sin(\alpha_1-\alpha_2)|\,(V_{11}+W_{12})
  \nonumber\\
  &=& 2|g_1g_2\sin(\alpha_1-\alpha_2)|\times \,{1\over 32a^6}
   (1+a^2+b^2)^4 \big((a^2+b^2+1)^2-3a^2\big)\,,
\end{eqnarray}
which is positive for all values of $a$ and $b$. The extremum is reached for
$a=1,\ b=0$ corresponding to $V_0= |g_1g_2\sin(\alpha_1-\alpha_2)|$.
This is obviously a minimum in the $a,b$ variables, but, since at this
point $C_-<0$, a maximum in the $SU(1,1)$ scalars.
Note that the positivity of the extremum is due to $T_-$.
For values of $a$ and $b$
outside the circle (\ref{circle1}) the $SU(1,1)$ scalars have a minimum.
In the case  $g_1g_2\sin(\alpha_1-\alpha_2)\,>\,0$ the potential is
negative everywhere.

\item $g_1=g_2=0$. In this case $C_-=(g_3^2+g_4^2)V_{33}$ and
$\Delta$ are proportional to $V_{33}$, which vanishes for
$a=1,\ b=0$, and is positive elsewhere. The analysis of Section \ref{sugra}
is therefore valid for all $a$ and $b$, and gives a minimum in the $r,\ \varphi$
variables, except in the point $a=1,\ b=0$, where the potential
(\ref{potrphi}) and $\Delta$ vanish.
This case corresponds to two $SO(3)$ Yang-Mills multiplets coupled to ungauged
supergravity. The absolute minimum at  $a=1,\ b=0$ corresponds to vanishing
cosmological constant. It is not difficult to see that this is the only
extremum of the potential for this particular choice of the matter sector.

\item $g_2=g_4=0$ etc. In all cases where we take one $SO(3)$ from the
supergravity
($R,S=1,\ldots,6$) and one $SO(3)$ from the matter sector there
are regions in $a$ and $b$ with $\Delta<0$.

\end{itemize}

\subsection{$SO(3)^2\times SO(2,1)^2$\label{examplesnc}}

For the noncompact group $SO(2,1)$ we need to assign three values
of the indices $R,S,\ldots$ corresponding to different values
of the diagonal metric $\eta$. We choose the groups in the following
way:
\begin{equation}
\label{groupdiv2}
R,S,\ldots = \overbrace{\,1\ 2\ 7\,}^{i=1}\ \overbrace{\,4\ 5\ 6\,}^{i=2}\
\overbrace{\,3\ 8\ 9\,}^{i=3}\ \overbrace{\,10\ 11\ 12\,}^{i=4}\,.
\end{equation}
The groups labelled 1 and 3 therefore correspond to $SO(2,1)$.
The $SO(2,1)$ structure constants are chosen as
\begin{equation}
   f_{RS}{}^T = -\epsilon_{RSU}\eta^{UT}\,,\qquad
   f_{RST} = -\epsilon_{RST}\,.
\end{equation}
It should be noted that the form of $V_{ij}$ and $W_{ij}$, and therefore
the resulting examples, depends crucially on the way the groups are distributed
over the 12 vector multiplets in (\ref{groupdiv2}). The analysis in this and the
following subsection is therefore far from exhaustive.

The contributions $V_{ij}$ and $W_{ij}$ to the potential now
take on the form:
\begin{eqnarray}
V_{11}&=&{1\over 16a^6}(a^2+b^2+1)^2 \big(b^2(a^2+b^2+1)^2 + 2a^2(a^2-b^2)\big)\,,
\nonumber\\
V_{22}&=& {1\over 64a^6}(a^2+b^2+1)^4\big((a^2+b^2+1)^2 - 6a^2\big)\,,
\nonumber\\
V_{33}&=&{1\over 16a^6}\big(b^2(a^2+b^2+1)^4 + 2a^2(a^2-b^2)(a^2+b^2+1)^2
           -8a^6\big)
\nonumber\\
V_{44}&=&{1\over 64a^6}\big((a^2+b^2+1)^2-4a^2\big)^2
    \big((a^2+b^2+1)^2+2a^2\big)\,,
\nonumber\\
V_{12}&=&V_{13}=V_{23}=V_{34}=0\,,
\nonumber\\
V_{24}&=& -{1\over 64a^6}\,(a^2+b^2+1)^3(a^2+b^2-1)^3\,,
\nonumber\\
W_{12}&=&W_{14}=W_{23}=W_{34}=0\,,
\nonumber\\
W_{13}&=&W_{24}=-{1\over 8a^6}b^3(a^2+b^2+1)^3\,.
\end{eqnarray}
In the search for interesting examples with just two nonzero coupling constants
$g_i$ and $g_j$ we find that
\begin{equation}
   \Delta = \big(2g_ig_j\sin(\alpha_i-\alpha_j)\big)^2\,
       \big(V_{ii}V_{jj}- V_{ij}^2\big)\,.
\end{equation}
Except for the case $i=2,\ j=4$, which concerns the two compact subgroups and
was already treated in Section \ref{examplesc}, the off-diagonal entries
in $V$ vanish, and the requirement $\Delta>0$ implies that for both subgroups
the corresponding $V_{ii}$ must be nonnegative. With the present choice
of subgroups this means that $i=2$ can be discarded. We find that both $V_{33}$
and $V_{44}$ are positive, except in the point $a=1,\ b=0$, where they vanish.
$V_{11}$ is positive everywhere, with a minimum at $a=1,\ b=0$ with
value $\half$.
Since also $W_{ij}$=0, either everywhere or at the point $a=1,\ b=0$, the
second term in the potential, $T_-$,
vanishes and will not help in obtaining a positive
cosmological constant.

The result with two groups will therefore always be such that the
extremum of the potential is at $a=1,\ b=0$, and that the potential
vanishes there.

\subsection{$SO(2,1)^4$\label{examplesnn}}

The structure constants are as in the Section \ref{examplesnc}.
The groups are now assigned in the following way:
\begin{equation}
R,S,\ldots = \overbrace{\,1\ 2\ 7\,}^{i=1}\ \overbrace{\,4\ 5\ 10\,}^{i=2}\
\overbrace{\,3\ 8\ 9\,}^{i=3}\ \overbrace{\,6\ 11\ 12\,}^{i=4}\,.
\end{equation}
The results for the potential do depend on how the groups are assigned,
clearly different choices are possible. The results with the choice above
are:
\begin{eqnarray}
V_{11}&=&V_{22}={1\over 16a^6}(a^2+b^2+1)^2
   \big(b^2(a^2+b^2+1)^2 + 2a^2(a^2-b^2)\big)\,,
\nonumber\\
V_{33}&=&V_{44}={1\over 16a^6}\big(b^2(a^2+b^2+1)^4 + 2a^2(a^2-b^2)(a^2+b^2+1)^2
           -8a^6\big) \,,
\nonumber\\
V_{ij}&=&0\ {\rm for}\ i\ne j \,,
\nonumber\\
W_{12}&=&W_{34}=0\,,
\nonumber\\
W_{13}&=&W_{24}=-{1\over 8a^6}b^3(a^2+b^2+1)^3\,,
\nonumber\\
W_{14}&=&-W_{23}=-{1\over 16a^6}b^2(a^2+b^2-1)(a^2+b^2+1)^3 \,.
\end{eqnarray}
The interesting case here is
\begin{itemize}
\item $g_3=g_4=0$. Now we have $C_-=V_{11}(g_1^2+g_2^2)$, which is
always positive, and
\begin{equation}
   V = V_{11}\,2|g_1g_2\sin(\alpha_1-\alpha_2)|\,.
\end{equation}
The potential is everywhere positive, the condition $\Delta>0$ is
trivially satisfied, and $C_->0$. There is therefore a absolute
minimum in all variables $r,\ \varphi,\ a$ and $b$. In this case there is no
contribution from $T_-$, and therefore (\ref{Vrel}) will be satisfied.
We will come back to further properties of this example in
Section \ref{discussion}
\end{itemize}

\section{Implications and discussion\label{discussion}}

We have analysed some special cases of gauged $N=4$ supergravity with six
additional matter multiplets with the aim of finding solutions
with positive cosmological constant. Two situations arise in which
this might happen: one is the example of $SO(3)^4$, where
a saddle point is found for a positive value of the potential: the
extremum is a minimum in the two matter fields considered, but a maximum
in the two $SU(1,1)$ scalars from the supergravity sector. This is an
extension with matter multiplets of the potentials studied by \cite{GZZ}.

Another
case involves $SO(2,1)^2$, where we find a positive minimum of the
potential in all variables. However, in this case the gauge group
is noncompact, which might give rise to wrong-sign kinetic terms.
The kinetic term of the $SU(1,1)$ scalars was given in (\ref{actionphi})
and does not have this problem.
Also the kinetic term of the matter scalars is independent of the
gauging. For the fields $a$ and $b$ one finds
\begin{eqnarray}
   {\cal L}_{{\rm kin}\,a,b} &=&
    -{3\over 4 a^4}\bigg( 4 a^2(1+b^2)\, (\partial a)^2 +
       ((a^2+b^2+1)^2 -4a^2b^2) \,(\partial b)^2 \nonumber\\
    &&\qquad +4ab(-a^2+b^2+1)\,\partial a\partial b \bigg) \,.
\end{eqnarray}
After diagonalizing this to new fields $a'$ and $b'$ one obtains
\begin{eqnarray}
    {\cal L}_{{\rm kin}\,a,b} &=&
    -{3\over 4 a^4}\bigg( 4a^2 (\partial a')^2
      + (1+a^2+b^2)^2 (\partial b')^2 \bigg)\,,
\end{eqnarray}
so these kinetic terms have indeed the standard sign for all $a,b$.
In the extremum $a=1,\ b=0$ one finds
\begin{equation}
   {\cal L}_{{\rm kin}\,a,b}\to -3 ( (\partial a)^2 + (\partial b)^2) \,.
\end{equation}
The potential and its second derivatives for this example are
in the extremum
\begin{equation}
   V(a,b) = |g_1g_2\sin(\alpha_1-\alpha_2)|(1
     + (a-1)^2 + 2 b^2 + \ldots ) \,.
\end{equation}
For these scalars the relation between the potential and its second derivatives
is not satisfied, since after a rescaling of $a$ and $b$ to get the
standard normalisation of the kinetic terms the (mass)${}^2$
of $a$ and $b$
still differ by a factor 2, and are not equal to $V_0$.

The vector kinetic terms do depend on the gauging,
and this might cause a problem for noncompact groups.
The vector kinetic terms are  for $a=1,\ b=0$,
\begin{equation}
\label{veckin}
   -\quart \eta_{RS} F^+_{\mu\nu}{}^RF^{\mu\nu\,+\,S}
   \bigg( {1\over \Phi_{(R)}}(\phi^1_{(R)} - \phi^2_{(R)}) -
        {2\over |\Phi_{(R)}|^2} \bigg) + {\rm h.c.}\,.
\end{equation}
In the case the gauging is $SO(3)^2$, as in the first example of Section
\ref{examplesc}, this is the form of the kinetic terms at the extremum of the
matter fields. The sum over $R,S$ then runs over the indices
$1,\ldots,6$, and the kinetic terms have the standard signs, as was shown
in detail
in \cite{MdRPW1}.
In the example of Section \ref{examplesnn} the form of the kinetic terms in
the matter extremum is again (\ref{veckin}), but now the indices run over the
values $1,2,7,4,5,10$ and two of the kinetic terms change sign.
This is a problem for the solution of Section \ref{examplesnn}.

In \cite{KLPS} it was found that potential and its second
derivatives in extended supergravity theories satisfies (\ref{Vrel}).
In Section \ref{sugra} we clarified this relation for the $SU(1,1)$
scalars. We have shown that (\ref{Vrel}) is modified by the term
$T_-$ in the scalar potential, but that for $T_-=0$ the relation is
indeed valid. As we have seen in this section, the matter scalars may violate
(\ref{Vrel}).
The relation (\ref{Vrel})
is reminiscent of supersymmetry Ward identities used in the past
to investigate cases with partially broken supersymmetry \cite{CGP}.
It would be interesting to make (\ref{Vrel}) more
precise in this way.

Much work remains to be done to give a more complete analysis of the
matter sector with six or more multiplets, and of
the examples that we have presented in this paper. Also their
possible application in cosmological models remains to be studied.
This is not only true for the example with an absolute minimum, but
equally so for cases with tachyonic modes.

\bigskip

\acknowledgments

\bigskip

It is a pleasure to thank Ulf Gran and
 Peter Wagemans for useful discussions.
The work of DBW is part of the research
programme of the ``Stichting
voor Fundamenteel Onderzoek van de Materie'' (FOM).
This work is supported in part by the European
Commission RTN programme HPRN-CT-2000-00131, in which MdR and DBW
are associated to the University of Utrecht.
SP thanks the University of Groningen for hospitality.


\newpage

\begin{thebibliography}{99}
\bibitem{AFV}R.~D'Auria, S.~Ferrara and S.~Vaul\`a,
{\sl $N=4$ gauged supergravity and a IIB orientifold with fluxes},
New.~J.~Phys. 4 (2002) 71,
{\tt hep-th/0206241}
\bibitem{AFLV}R.~D'Auria, S.~Ferrara, M.A.~Lled\'o and S.~Vaul\`a,
{\sl No-scale $N=4$ supergravity coupled to Yang-Mills: the
scalar potential and super-Higgs effect},
{\tt hep-th/0211027}
\bibitem{DFR}A.~Das, M.~Fischler and M.~Ro\v{c}ek,
{\sl Super-Higgs effect in a new class of scalar models and a model of
super QED},
Phys.~Rev.~D16 (1977) 3427
\bibitem{FS}D.~Z.~Freedman and J.~H.~Schwarz,
{\sl $N=4$ supergravity theory with local $SU(2)\times SU(2)$
  invariance},
Nucl.~Phys.~B137 (1978) 333
\bibitem{GZZ}S.J.~Gates and B.~Zwiebach,
{\sl Gauged $N=4$ supergravity theory with a new scalar potential},
Phys.~Lett.~123B (1983) 200;\\
B.~Zwiebach,
{\sl Gauged $N=4$ supergravity and Gauge Coupling Constants},
Nucl.~Phys.~B238 (1984) 367
\bibitem{MdR1}M.~de Roo, {\sl Matter coupling in $N=4$ supergravity},
Nucl.~Phys.~B255 (1985) 515
\bibitem{MdRPW1}M.~de Roo and P.~Wagemans,
{\sl Gauged matter coupling in $N=4$ supergravity},
Nucl.~Phys.~B262 (1985) 644
\bibitem{PW1}P.~Wagemans,
{\sl Aspects of $N=4$ supergravity},
Groningen, thesis, 1990
\bibitem{GWG}G.~W.~Gibbons, {\sl Aspects of supergravity theories},
in {\sl Supersymmetry, Supergravity and Related Topics},
editors F.~Del Aguila, J.~A.~de Azc\'arraga and L.~E.~Ib\'a\~{n}ez,
World Scientific 1985
\bibitem{dWHDS}B.~de Wit, D.J.~Smit and N.D.~Hari Dass,
{\sl Residual supersymmetry of compactified $D=10$ supergravity},
Nucl.~Phys.~B283 (1987) 165
\bibitem{MN}J.~Maldacena and C.~Nunez,
{\sl Supergravity description of field theories on curved manifolds and
a no go theorem},
Int.~J.~Mod.~Phys.~A16 (2001) 822,
{\tt hep-th/0007018}
\bibitem{FTVP}P.~Fr\'e, M.~Trigiante and A.~Van Proeyen,
{\sl Stable de Sitter Vacua from $N=2$ Supergravity},
Class.~Qu.~Grav.~19 (2002) 4167,
{\tt hep-th/0205119}
\bibitem{KLPS}R.~Kallosh, A.~Linde, S.~Prokushkin and M.~Shmakova,
{\sl Gauged Supergravities, de Sitter Space and Cosmology},
Phys.~Rev.~D65 (2002) 105016
{\tt hep-th/0110089}
\bibitem{BMdR}H.~J.~Boonstra and M.~de Roo,
{\sl Duality symmetry in four-dimensional string actions}
Phys.~Lett.~B353 (1995) 57,
{\tt hep-th/9503038}
\bibitem{MS}J.~Maharana and J.~H.~Schwarz,
{\sl Noncompact symmetries in string theory},
Nucl.~Phys.~B390 (1993) 3,
{\tt hep-th/9209052}
\bibitem{CGP}S.~Cecotti, L.~Girardello and M.~Porrati,
{\sl Constraints on partial super-Higgs},
Nucl.~Phys.~B268 (1986) 295




\end{thebibliography}
\end{document}